\documentclass[iop,apjl]{emulateapj}
\usepackage{ulem}
\usepackage{apjfonts}
\usepackage{graphicx}
\usepackage{epsfig}
\usepackage{amssymb}
\usepackage{amsbsy}
\usepackage{amsmath}
\usepackage{amsfonts}
\usepackage{mathrsfs}
\usepackage[below]{placeins}
\usepackage{natbib}
\usepackage{multirow}
\usepackage{enumerate}
\usepackage{color}
\usepackage[plainpages=false,colorlinks=true,anchorcolor=red,linkcolor=blue,citecolor=black,urlcolor=blue,bookmarks=false]{hyperref}
\def\gtorder{\mathrel{\raise.3ex\hbox{$>$}\mkern-14mu
    \lower0.6ex\hbox{$\sim$}}}
\def\ltorder{\mathrel{\raise.3ex\hbox{$<$}\mkern-14mu
    \lower0.6ex\hbox{$\sim$}}}

\def\msun{~\mathrm{M}_\odot}
\def\hmsun{~h^{-1} \mathrm{M}_\odot}

\def\hmpc{~h^{-1} \mathrm{Mpc}}
\def\hkpc{~h^{-1} \mathrm{kpc}}
\def\mgal{M_{\rm gal}}

\newcommand{\code}[1]{\texttt{#1}}

\slugcomment{To be published by the Astrophysical Journal Letters}
\shortauthors{Romano-D\'{\i}az, Shlosman, Choi and Sadoun}
\shorttitle{The Growth of galaxies at high redshifts}

\begin{document}

\title{The Gentle growth of galaxies at high redshifts in overdense environments}

\author{Emilio Romano-D\'{\i}az$^1$} 
\author{Isaac Shlosman$^{2,3}$} 
\author{Jun-Hwan Choi$^4$}  
\author{Raphael Sadoun$^5$}

\affil{$^1$
Argelander Institut fuer Astronomie, University of Bonn, Auf dem Haegel 71, D-53121 Bonn, Germany}
\affil{$^2$
Department of Physics and Astronomy, University of Kentucky, Lexington, KY 40506-0055, USA}
\affil{$^3$ 
Theoretical Astrophysics, Department of Earth \& Space Science, Osaka University, Osaka 560-0043, Japan}
\affil{$^4$
Department of Astronomy, University of Texas, Austin, TX 78712-1205, USA}
\affil{$^5$
Deptartment of Physics \& Astronomy, University of Utah, Salt Lake City, UT 84112-0830, USA}

\begin{abstract}
We have explored prevailing modes of galaxy growth for redshifts $z\sim 6-14$, comparing 
substantially
overdense and normal regions of the universe, using high-resolution zoom-in cosmological
simulations. Such rare overdense regions have been projected to host high-$z$ quasars. We
demonstrate that galaxies in such environments grow predominantly by a smooth accretion from 
cosmological filaments which dominates the mass input from major, intermediate and minor mergers. 
We find that by $z\sim 6$, the accumulated galaxy mass fraction from mergers falls short by a 
factor of 10 of the cumulative accretion mass for galaxies in the
overdense regions, and by a factor of 5 in the normal environments. Moreover, the 
rate of the stellar mass 
input from mergers also lies below that of an {\it in-situ} star formation (SF) rate.
The fraction of stellar masses in galaxies contributed by mergers in overdense regions
is $\sim 12\%$, and $\sim 33\%$ in the normal regions,
at these redshifts. Our median SF rates for $\sim {\rm few}\times 10^9\msun$
galaxies agrees well with the recently estimated rates for $z\sim 7$ galaxies
from Spitzer's SURF-UP survey. Finally, we find that the main
difference between the normal and overdense regions lies in the amplified
growth of massive galaxies in massive dark matter halos.
This leads to the formation of $\gtorder 10^{10}\msun$ galaxies due to the
$\sim 100$-fold increase in mass during the above time period. Such galaxies are basically absent 
in the normal regions at these redshifts.   
\end{abstract}



\section{Introduction}
\label{sec:intro}

Detection of luminous quasars with supermassive black holes of $\sim
10^9\msun$ at $z\gtorder 6$
\citep[e.g.,][]{Fan2003,Willott2010,Mortlock2011} have raised a long
list of issues, including properties of parent galaxies, their host
dark matter (DM) halos, and the environment.  Contradictory claims
have been made on whether quasars reside in overdense and rare regions
of the universe \citep[e.g.,][]{Stiavelli2005,Willott2005,Zheng2006,
  Kim2009,Overzier2009,erd11a}. The properties of high-$z$ quasar-host
galaxies are virtually unknown.  In this Letter we compare galaxy
evolution at high redshifts in such highly overdense regions with
those in `normal' regions, using high-resolution numerical
simulations. Specifically, we ask what is the main growth mode(s) of
galaxies in these environments, and quantify the rates of (1) smooth
accretion from cosmological filaments, and of (2) galaxy mergers.

With the advent of sensitive near-infrared imaging on the HST and
JWST, and the new generation of ground-based telescopes, more than 100
galaxies have been detected at $z\sim 6.5 - 9$
\citep[e.g.,][]{Finkelstein2010,Mclure2011,Bouwens2011,Lorenzoni2011,McCracken2012,
Trenti2012,Matthee2014,Ryan2014}, and their star formation (SF) rates and
mass and luminosity functions analyzed 
\citep[e.g.,][]{Stiavelli2004,Yan2004,Finlator2011,erd-I,erd11b,Jaacks2012,
Yajima2012,Yajima2014,Dayal2013}. The $\sim 3\sigma$ overdensities
of Lyman Break galaxies (LBGs) --- probably the most massive galaxies
at these epochs \citep[e.g.,][]{Baugh98,Nagamine04}, have been
revealed on scales of a few Mpc around high-$z$ quasars, using
Suprime-Cam of Subaru telescope \citep[e.g.,][]{Utsumi2010}.

The new paradigm of galaxy growth by accretion from cosmological
filaments has succesfully challenged the merger-dominated scenario
\citep[e.g.,][for
review]{Kerevs2005,Kerevs2009,Dekel2006,Dekel2009,Finlator2011,Shlosman2013}.
Numerical simulations indicate that major mergers are not the main
mode of growth, nor are they responsible for the intense SF
\citep[e.g.,][]{Dayal2013}, although the numerical resolution is still
low. Moreover, simulations hint at a bimodal behavior with a cold gas
accretion dominating in low-mass galaxies and the hot gas in the
higher-mass ones \citep[e.g.,][]{Kerevs2009} ---
in tandem with the observed trend in SDSS galaxies
\citep[e.g.,][]{Kauffmann2004}.

Motivated by these issues, we study the main modes of galaxy growth,
in gas and stellar components, at $z\sim 6-14$, to facilitate
further comparison with observations. Using the Constrained
Realizations method (section \ref{sec:ICs}) to sample rare overdense
regions without the loss of generality at a very high resolution, we
aim at quantifying mass accretion rates from cosmological filaments
and from galaxy mergers, as a function of the environment.

\section{Numerics and Initial Conditions}
\label{sec:ICs}

We use the modified tree-particle-mesh Smoothed
Particle Hydrodynamics (SPH) code \code{GADGET-3} \citep{gadget2}, in
its conservative entropy formulation \citep{Springel2002}. Our
implementation includes radiative cooling by H, He and metals
\citep{Choi2009}, SF, stellar feedback, a phenomenological model for
galactic winds, and sub-resolution model for multiphase interstellar
medium \citep[ISM;][]{Springel2003}, where starforming SPH particles
contain cold and hot phases. The SF prescription is based on the
''Pressure model'' which reduces the high-$z$ SF rate
\citep{choi-nagamine10} relative to previous implementations. SF can
only take place with gas density above the threshold $n_{\rm H,SF} =
0.6~\mathrm{cm}^{-3}$.

We have used the Constrained Realizations [CR] of \citet{erd11b} to
obtain initial conditions [ICs], as well as their unconstrained
counterpart [UCR] for comparison. The CRs have been implemented
following the algorithm of \cite{hr91} \citep[see also][]{erd07} and
are similar to those in \cite{erd11b}.  The constraints were imposed
onto a grid of 1,024 cells per dimension in a cubic box of $20\hmpc$,
to prescribe the formation and collapse of a massive halo of $M_{\rm
  h}=10^{12}\hmsun$ by $z \sim 6$, according to the top-hat
model. $h=0.701$ is the Hubble constant in units of
$100\,\mathrm{km\,s^{-1} Mpc^{-1}}$. We assumed
$\Lambda$CDM-WMAP5 cosmological parameters \citep{Dunkley2009},
$\Omega_{\rm m} = 0.28, \Omega_{\Lambda}=0.72, \Omega_{\rm b} =
0.045$.  The power spectrum is normalized by the linear rms amplitude
of mass fluctuations in $8\hmpc$ spheres, $\sigma_8 = 0.817$. For the
UCR simulation, we have used a factor of two smaller resolution
($2\times 512^3$).

The simulations have been followed from $z=199$ to $z\sim 6$ with a
time sampling of $\Delta t = 10$\,Myr between output-times. We have
applied the zoom-in technique with three levels of refinement to
increase the mass resolution, with baryons only at the highest
refinement level. For the CR simulation the inner region has a radius
of $3.5\hmpc$, with an effective resolution of $1024^3$ in both DM and
SPH particles, and a gravitational softening of $\epsilon_{\rm grav} =
300$\,pc (comoving). The corresponding particles masses are: $4.66
\times 10^5\hmsun$ (DM), and $1.11 \times 10^5\hmsun$ (gas).  Gas
particles experience up to two episodes of SF, resulting in stellar
masses of $5.55 \times 10^4\hmsun$. For the UCR, the inner region has
a radius of $7\hmpc$, and mass resolutions of $3.73\times 10^6\hmsun$
(DM), $8.9\times 10^5\hmsun$ (gas), and $4.4\times 10^5\hmsun$
(stars).

Galaxies are identified by means of the HOP group finder algorithm
\citep{hop}, for densities exceeding $0.01 n_{\rm H,SF}$. 
We consider only objects with the
minimal {\it stellar} mass above $\mgal = 2.8\times 10^7\hmsun$
\citep[in agreement with recent observations,][]{Ryan2014},
independently of their gas content, i.e., 
$N \ge 500$ stellar particles for the CR and 65 for UCR models.  This
ensures that the corresponding DM-parent halos are $\gtorder
10^8\msun$, and therefore, are almost unaffected by reionization
\citep[e.g.,][]{Barkana1999}.

We analyze spherical volumes of radii $r_{\rm CR} = 2.7\hmpc$ and
$r_{\rm UCR} = 4.0\hmpc$ at $z=6.2$, free of contamination from
massive DM particles. The UCR volume has been matched to contain the
same amount of matter as in the CR.  For each galaxy, we construct its
baryonic mass accretion history (MAH), until its progenitor (parent
galaxy) cannot be further identified. A progenitor is defined as an
object that provides at least 51\% of the mass with respect to the
child galaxy.

\begin{figure*}[ht!!!!!!!!!!!!!]
  \begin{center}
    \includegraphics[angle=0,scale=0.7]{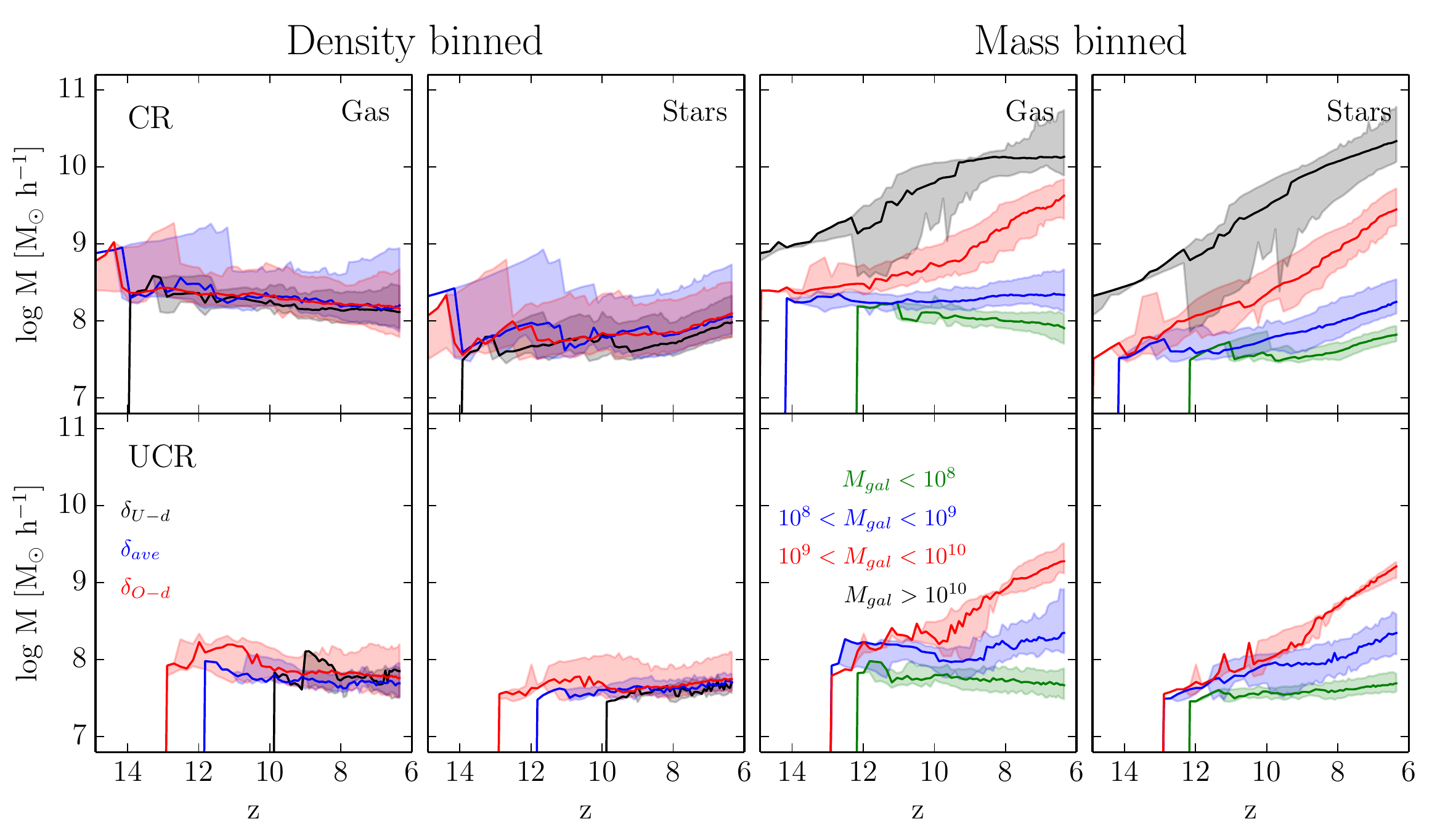}
  \end{center}
  \caption{Mass accretion histories (MAHs) of the gaseous and stellar
    components in galaxies of the CR-simulation (top row) and
    UCR-simulation (bottom row). The samples are binned by their local
    density (left panels) and stellar masses (right panels). The
    continuous lines represent the median of the distributions while
    the shaded regions give the 20 (lower boundary) and 80 (upper
    boundary) percentiles. The masses on the y-axes correspond to the
    gas and stellar masses in galaxies, respectively.  }
\label{fig:mahs}
\end{figure*}

From this sample, we choose galaxies whose MAHs could be
followed beyond $z=8$, giving us enough span-time to
analyze their growth. The final sample is composed of
$N_{\rm CR} = 271$ and
$N_{\rm UCR} = 97$ objects.  The galaxy masses range $M_{\rm gal}\sim
[2.8\times 10^7 - 1.2\times 10^{11}]\hmsun$. The halos can in principle
host more than one galaxy. We do not distinguish between main and
satellite galaxies, giving them equal treatment.

The galaxy growth is split into smooth accretion and mergers.
For each output time we identify all progenitors that will
merge within $\Delta t$ with the main branch of the chosen galaxy
and divide those merger events into: 
(1) Major merger: with a mass ratio of 1-to-3,  
(2) Intermediate: with a mass ratio of 4-to-10,  
(3) Minor: with a mass ratio of 11 and above.
For stellar accretion, we also identify {\it in-situ} SF, as those 
stars that are  born within $\Delta t$.

Statistical analysis of galaxy properties is carried out by binning
galaxies in two different ways: (1) according to their stellar masses
($\mgal$), and (2) according to their relative smoothed density
contrast, $\delta = (\rho_{\rm bar} - \tilde\rho_{\rm
  bar})/\tilde\rho_{\rm bar}$), with respect to the mean baryon
density $\tilde\rho_{\rm bar}$ (when convolved with a top-hat filter
of $250\hkpc$) within the simulated baryonic volume. Therefore,
galaxies are divided into (1) Underdense regions: $\delta < 0$, (2)
Average-density regions: $\delta = [0,3]$, and (3) Overdense regions:
$\delta > 3$. The CR simulation describes the highly overdense region,
$\sim 5\sigma$ with respect to the average density of the universe,
while the UCR simulation has the average density. Our results are not
sensitive to the size of the kernel as long as it is larger than the
average galaxy inter-separation within the volume.

\section{Results}
\label{sec:results}

Our goal is to compare the galaxy growth in various environments and
estimate contributions from mergers and smooth accretion.
Figure~\ref{fig:mahs} shows the median of the MAH distributions for
the CR and UC simulations, averaged over periods of 50\,Myr.
We choose the median over the mean
because the distributions are not Gaussian but substantially
skewed, and the mean is particularly sensitive to the influence
of outliers.
The first and second columns show the distributions for the
underdense, average and overdense regions (density-binned case). The
third and last columns represent the mass-binned distributions.

The CR and UCR distributions differ substantially. The density-binned
distributions show a rather flat behavior due to the mixing of various
masses within a given region, with a slightly decreasing trend for the
gaseous component and increasing one for the stars in CR. The UCR is
even flatter and has a lower median mass by a factor of $\sim 2$.  As
dicussed below, the most massive galaxies are virtually absent from
the low density regions.

On the other hand, the mass-binned case shows a clear growth trend
among the various mass bins, especially for higher masses. Two are the
main differences that strive between the CR and UCR simulations due to
the effect of the imposed constraint (section\,\ref{sec:ICs}).  First,
the CR galaxies with $\mgal > 10^8\hmsun$ form earlier than in the UCR
case, which can also be seen for the density-binned curves. Although
different resolutions could play a role, mostly in triggering the
onset of SF, this does not affect the general results --- the least
massive objects, those with $\mgal < 10^8\hmsun$, exhibit the same
distribution and the time appearance in CR and UCR runs.

Secondly and most importantly, massive galaxies with $\mgal >
10^{10}\hmsun$ in the UCR simulation are completely absent, since no
halos with masses above $10^{11}\hmsun$ form before $z\sim 6$. At
large, the growth trends for both gas and stars are similar in both
simulations (once the SF takes place in the UCR simulation), with one
important difference --- the mass accumulation in galaxies in the
overdense regions proceeds at an accelerated rate.  Galaxies with
$\mgal > 10^{9}\hmsun$ show that the median mass increases as
log\,$\mgal\sim z$, i.e., $\mgal$ increases exponentially (with the
exception of gas mass which saturates at $z\sim 9.5$ in $\mgal >
10^{10}\hmsun$, due to the SF).

The stellar and gaseous components in galaxies with $\mgal >
10^{9}\hmsun$ grow faster than in less massive galaxies. These massive
objects are mostly found in the over- and average-density regions,
favorable to the mass accumulation by smooth accretion and mergers.

\subsection{Gas Accretion}
\label{sec:gas}

\begin{figure*}[ht!!!!!!!!!!!]
  \begin{center}
    \includegraphics[angle=0,scale=0.75]{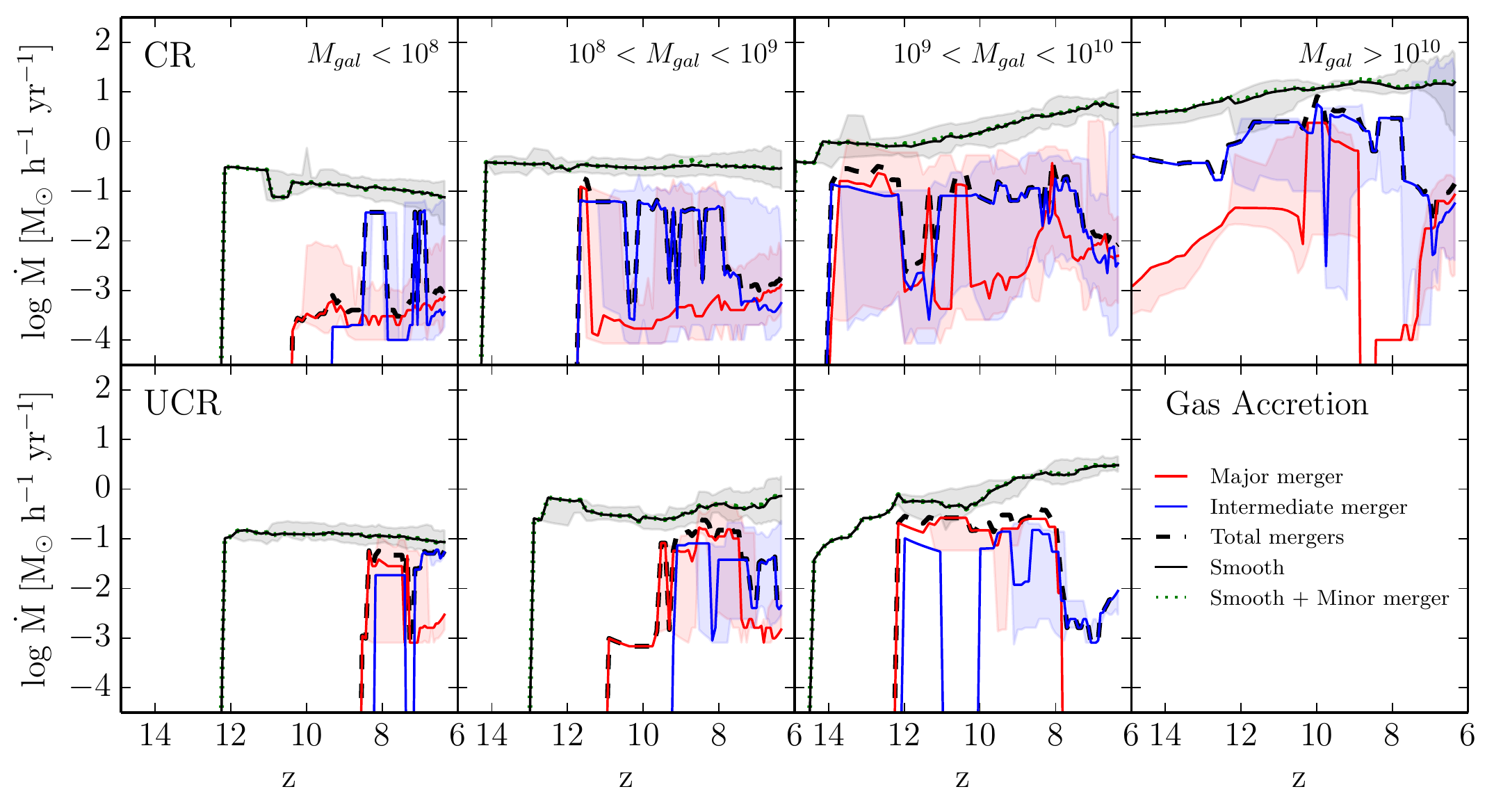}
  \end{center}
  \caption{Gas accretion rate histories for the CR simulation (top
    row) and UCR simulation (bottom row) for the mass-binned
    distributions (as in Fig.~\ref{fig:mahs}). The contributions are
    split into major mergers (red), intermediate mergers (blue),
    smooth accretion (black) and smooth plus minor mergers (green
    dotted line). The total merger contribution (major$+$intermediate)
    is shown by the thick-dashed black line.}
\label{fig:gas}
\end{figure*}

Figure~\ref{fig:gas} shows the mass-binned gas accretion rates for the
CR and UCR simulations.  Because the lower mass cutoff in minor
mergers depends on the numerical resolution limit, and because their
overall contribution is small, we have added them to the smooth
accretion.  As before, the trends provide the median of the
distributions.

The mass-binned distributions clearly show that the smooth accretion
rate dominates over the total merger contribution at all $z$, in all
mass bins and for both simulations. Furthermore, minor merger mass
input rate appears to be insignificant when compared to the smooth
accretion rate, as shown by the green-dotted lines which are almost
identical to the the black lines representing smooth accretion.  For
the CR run, intermediate merger rate is more than an order of
magnitude below the smooth accretion, and even its peaks reach only
$\approx 40\%$.

The major merger input rate is naturally more spiky, but generally
falls well below that of the intermediate ones. For CR galaxies,
mergers play some role only for $z > 10$ (and only for $10^9 -
10^{10}\hmsun$ bin), and an intermittent role at lower $z$. However,
their spiky median distributions and shadowed regions, which could
overlap with the smooth accretion trends, indicate that sometimes
their contribution can be substantial. Similarly for the UCR
simulation --- major mergers dominate among mergers at high-$z$
(although lower than CR due to the delayed evolution of the normal
regions with respect to the overdense ones), and intermediate mergers
contribute more at $z < 8$.  Galaxies with $\mgal < 10^9\hmsun$
exhibit a slow constant growthrate, while more massive objects show an
accelerated exponential growth as
discussed above.  As these massive galaxies prefer an overdense
environment, where the number density of potential mergers as well as
of the background IGM is highly elevated, we observe this growthrate
bifurcation.  There is an overall decline of merger gas input rate
after $z\sim 8$ for massive galaxies because their gas fraction
decreases with time.

Overall, Figure~\ref{fig:gas} shows that the gas supply rate to galaxies is dominated
by the smooth accretion over the mergers. Among mergers,
the intermediate ones dominate the major mergers, as seen from the thick dashed
lane which nearly always coincides with the solid blue lane.

\subsection{Stellar Accretion and Star Formation}
\label{sec:stellar}

\begin{figure*}[ht!!!!!!!!]
  \begin{center}
    \includegraphics[angle=0,scale=0.75]{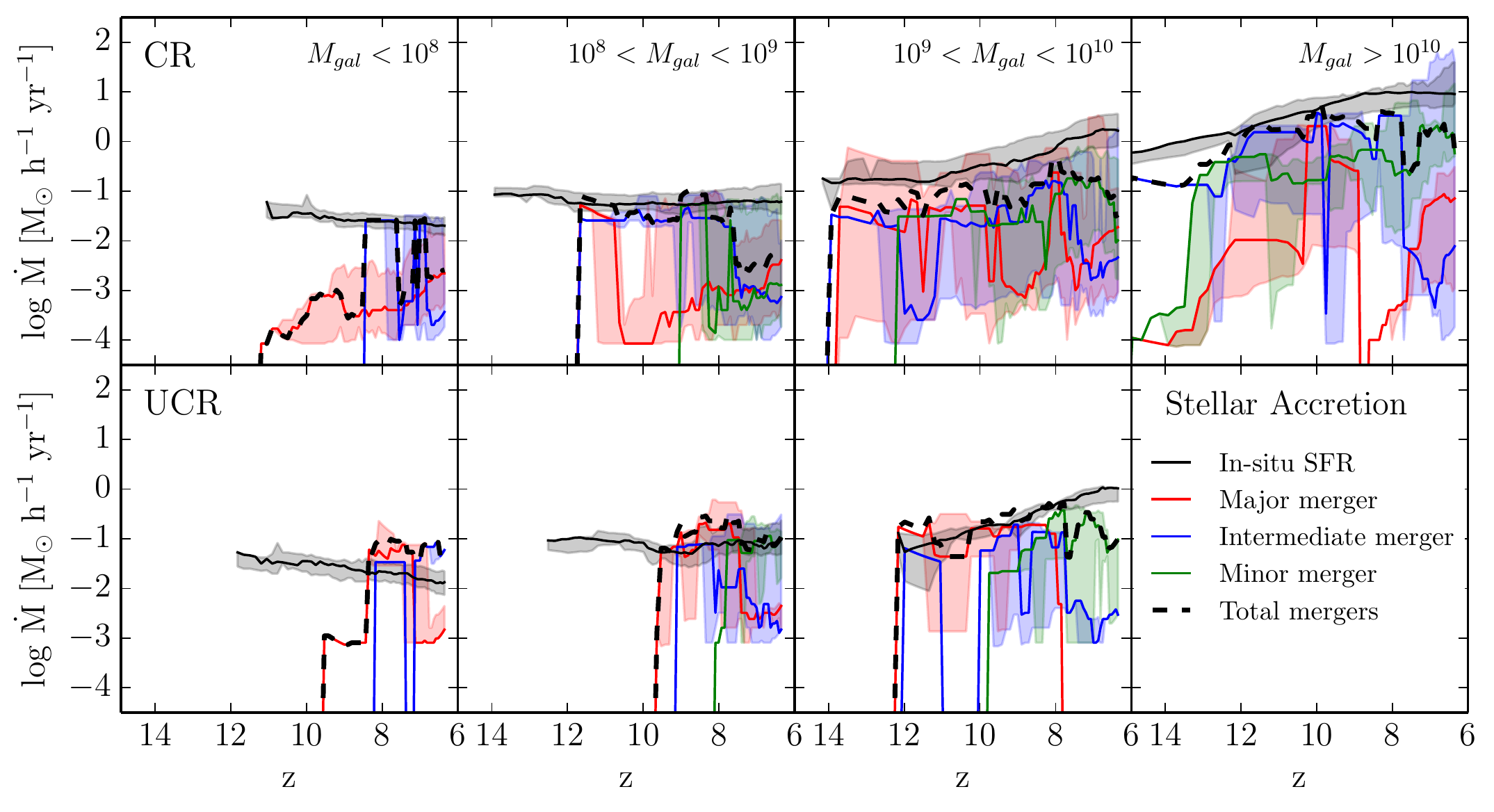}
  \end{center}
  \caption{Same as Figure~\ref{fig:gas} but for the stellar mass
    evolution rate. The black line represents the {\it in-situ} SF
    rate and there is no contribution from the smooth accretion.}
\label{fig:stellar}
\end{figure*}

The behavior of stellar accretion mirrors that of the gas in many
aspects but differs because of the {\it in-situ} SF.  The SF rate
dominates the stellar mass growth in galaxies for all mass-bins,
independently of environment (Fig.~\ref{fig:stellar}).  Here we show
explicitly the minor merger contributions since there is no stellar
counterpart of the smooth gas accretion.  The SF rate successfully
competes with the contributions from mergers in the CR and UCR
simulations. For limited time periods of merger activity in the UCR
simulation, the mergers can dominate.

The SF rate exhibits similar exponential growthrate with $z$ for
$\mgal > 10^9\hmsun$, and is flat for the lower masses, where major
merger activity dominates.  The similarity in the SF rate to that of
the smooth gas accretion serves an indication that most of the
accreted gas is cold and metal-rich, thus able to facilitate the SF,
as elaborated elsewhere.

\begin{figure*}[ht!!!!!!!!]
  \begin{center}
    \includegraphics[angle=0,scale=0.75]{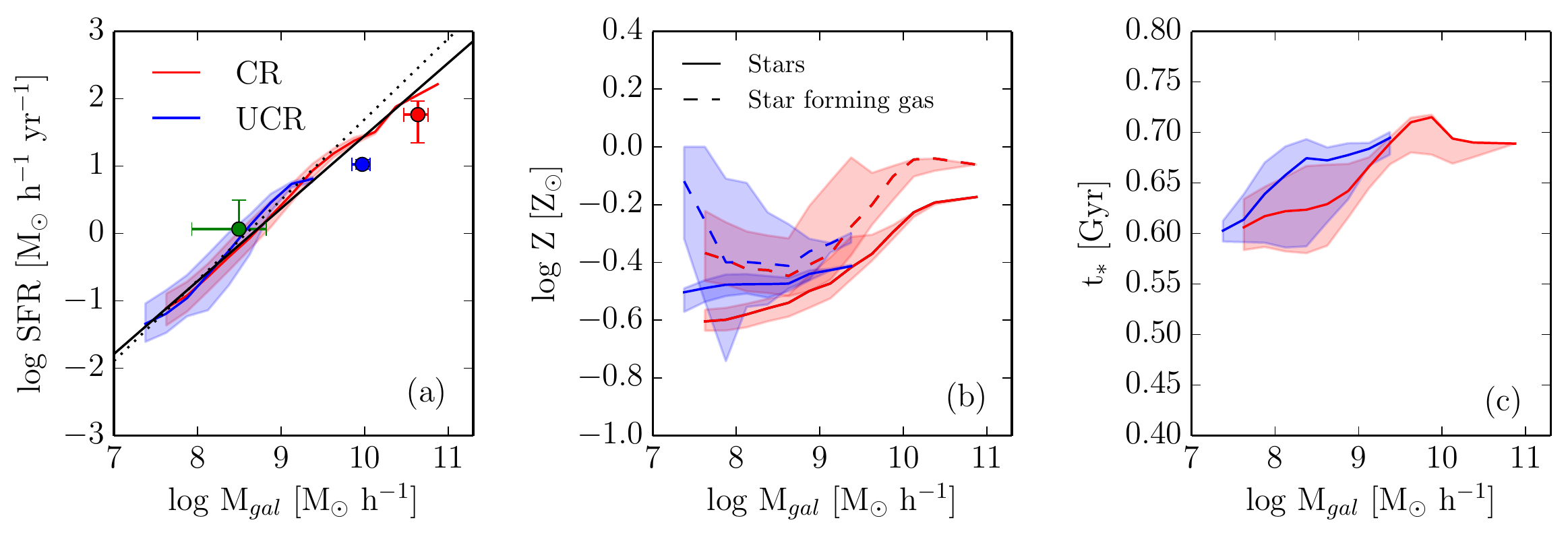}
  \end{center}
  \caption{Galaxy properties in CR and UCR simulations as a function
    of $M_{\rm gal}$ at $z\sim 6$. {\it (a)}: the SFR --- the black
    solid and dotted lines are the least-square fits: log\,$SFR
    [M_\odot\,{\rm yr^{-1}}]\approx 1.19\,{\rm log}\,(M_{\rm
      gal}/M_\odot)$ (UCR) and log\,$SFR [M_\odot\,{\rm
      yr^{-1}}]\approx 1.08\,{\rm log}\,(M_{\rm gal}/M_\odot)$
    (CR). Observations: the red square \citep{Dow-Hygelund2005}, green 
    square \citep{Ono2010}, the blue pentagon \citep{Lai2007}; 
    {\it (b)}: metallicity $Z/Z_\odot$; {\it (c)}: the median
    stellar age $t_*$.  The continuous and dashed lines represent the
    median of the distributions, while the shaded regions give the 20
    (lower boundary) and 80 (upper boundary) percentiles.
 }
\label{fig:properties}
\end{figure*}

We also note that for the CR galaxies located in overdense
environment, $\delta > 3$, the overall merger mass input represents
more than $50\%$, mostly comprised of intermediate and minor mergers,
although major mergers do contribute more to the growth of the stellar
content than in the case of the gas.

The fraction of the {\it total} mass accumulated by the mergers is
$\sim 9\%$ (CR) and $\sim 20\%$ (UCR) for $\mgal\ltorder
10^{10}\hmsun$. The fraction of stars in galaxies contributed by
mergers is $\sim 12\%$ (CR) and $\sim 33\%$ (UCR).

By analyzing
the physical properties of galaxies in our simulations, we have
determined that the tight correlations SFR\,$-M_{\rm gal}$ and and
metallicity $Z-M_{\rm gal}$ of GOODS LBGs
\citep{Finlator2006,Finlator2011} for the UCR can be extended to
$z\gtorder 6$ and to large overdensities (CR)
(Figs.~\ref{fig:properties}a-b), and show excellent agreement with 
few known observing points (Fig.~\ref{fig:properties}a).  We also observe 
that more massive galaxies have older median stellar populations
(Fig.~\ref{fig:properties}c) --- the anticipated behavior at high $z$
when the SF rate is still rising.

The UCR simulations can be directly compared to \citet{Dayal2013},
which are of lower resolution in mass and spatially. We find that
indeed the merger mass input rate drops towards $z\sim 6$, but only
for the gas. The stellar contribution from the mergers continues to
compete with the {\it in-situ} SF.  For the CR run, the relative
contribution of mergers to the gas and stars is lower than in the
UCR. Partly, this difference with \citet{Dayal2013} can be explained
by their definition of a galaxy which is based on the minimum of 10
stellar particles, compared to our 500.

\section{Discussion}
\label{sec:discussion}

We have used high-resolution zoom-in cosmological simulations to study
the growth of galaxies in highly overdense and normal environments at
$z\sim 6-14$. The initial conditions for the overdense field have been
obtained by means of the Constrained Realizations (CR) method which
allows to sample rare regions without the loss of generality, while no
constraints have been applied in the normal (UCR) region. Using the
same Gaussian random realization as the UCR run, the CR simulation box
contains the seed of a massive DM halo of $M_{\rm h}\sim
10^{12}\hmsun$ expected to collapse by $z\sim 6$.  We have compared
the galaxy MAHs for the CR and UCR simulations and have analyzed the
contributions from the smooth accretion, minor, intermediate and major
mergers to the growth of the gaseous and stellar components in
galaxies.

Galaxies have been defined in terms of their stellar mass.
The minimal resolved mass is $\mgal = 2.8\times 10^7\hmsun$
\citep[e.g.,][]{Ryan2014}.  For comparison, the most massive galaxy is
$\sim 1.2\times 10^{11}\msun$ situated in the $M_{\rm h}\sim
10^{12}\msun$ halo at $z\sim 6$.  Galaxies with $\mgal\gtorder
10^{10}\hmsun$ can represent high-$z$ LBGs detected at $z\sim 6-7$
\citep[e.g.,][]{Utsumi2010}. Note that the halo occupation number is
larger than unity.

Our main result concerns the prevailing mode of galaxy growth. The
overdense environment considered here is a typical one to obtain
collapse of $\sim 10^{12}\msun$ DM halo by $z\sim 6$ in the top-hat
model. The underlying DM structure of the region is responsible for
the accelerated galaxy evolution observed in the CR.  We find that the
smooth accretion rate from cosmological filaments substantially
dominates over that from mergers. Consequently, by $z\sim 6$, the
assembled median galaxy mass which has been contributed by mergers
(gas$+$stars) falls by more than an order of magnitude below the
accretion input for CR and by a factor of 5 for the UCR galaxies.

Second, we find that the growthrate of the stellar mass in mergers in
overdense regions lies also below the {\it in-situ} SF rate. The
accumulated stellar mass due to the merger contribution falls short of
the locally-produced stars, only $\sim 12\%$ of the total stellar mass
for the CR and $\sim 33\%$ for the UCR objects. Our median SF rates
for galaxies with $\mgal\sim {\rm few}\times 10^9\hmsun$ agrees nicely
with the current measure of $z\sim 7$ galaxies using Spitzer's
SURFS-UP survey \citep[][]{Ryan2014}.

Finally, we find that the main difference between the normal and
overdense regions lies in that galaxy evolution rate in the latter
ones has been amplified and leads to the formation of more massive
galaxies, which are absent in the normal regions.  Fig.~\ref{fig:mahs}
underlines this trend: in both CR and UCR runs, only galaxies $\mgal >
10^9\hmsun$ amplify their growth rate. This leads to the median
stellar mass of the CR galaxies in the highest bin to `fatten' by a
factor of $\sim 100$ during $z\sim 6-14$.

Merger activity becomes relevant at $z \ltorder 14$, especially in the
overdense regions and for the most massive galaxies.  The decrease in
the major merger activity in bringing in the gas indicates that the
relative fraction of the gas in galaxies becomes lower with $z$, which
is confirmed directly.

The gas accretion does depend on the type of feedback mechanisms
imposed in the numerical modeling. While in principle the AGN feedback
(neglected here) could halt the smooth accretion which proceeds mainly
along the filaments, this is rather inconclusive, as this feedback is
known to be anisotropic.
 
\section*{Acknowledgments}

We thank Volker Springel for providing us with
the original version of GADGET-3 and Yehuda Hoffman for preparing the
initial conditions using the Constrained
Realizations. E.R.D. acknowledges support from the SFB 956 by the DFG. 
I.S. acknowledges partial
support by the NSF, NASA and STScI.  Simulations have been
performed on the UK DLX Cluster.


\end{document}